\documentclass[12pt]{article}

\newcommand{\bbr}{I\!\! R}

\newcommand{\x}{arXiv:}
\newcommand{\m}{\mathrm}

\usepackage{graphicx}

\begin{document}
\thispagestyle{empty}
\begin{center}

\null \vskip-1truecm \vskip2truecm

{\Large{\bf \textsf{Bounding the Temperatures of Black Holes Dual to Strongly Coupled Field Theories on Flat Spacetime}}}

{\large{\bf \textsf{}}}

{\large{\bf \textsf{}}}

\vskip1truecm

{\large \textsf{Brett McInnes
}}

\vskip1truecm

\textsf{\\  National
  University of Singapore}

\textsf{email: matmcinn@nus.edu.sg}\\

\end{center}
\vskip1truecm \centerline{\textsf{ABSTRACT}} \baselineskip=15pt
\medskip

We show that AdS black holes dual to field theories on flat spacetime, as used in applications of
the AdS/CFT correspondence to strong interaction and condensed matter physics, have temperatures with positive
lower bounds. There are two distinct effects involved. For low chemical
potentials in the dual field theory, the cooling black hole makes a transition to a state corresponding to
confinement in the field theory. For high chemical potentials, it becomes unstable to
a non-perturbative string effect. This allows a holographic sketch of the field theory phase diagram, one
which is in qualitative agreement with the phenomenological understanding of the theory at [relatively] low temperatures.
It also puts an interesting upper bound
on the temperature-normalized chemical potential $\bar{\mu}$ of the field theory, if it describes a plasma: in the normalization of Myers et al., $\bar{\mu}$ must be less than
approximately 0.49. Thus, the extent to which a chemical potential can worsen violations of the KSS bound is severely restricted.

\newpage

\addtocounter{section}{1}
\section* {\large{\textsf{1. Can A Black Hole Be Arbitrarily Cold?}}}
Extremal black holes are important but puzzling objects: they have zero temperature
but non-zero entropy. Many [see for example \cite{kn:lemo} and its references to the earlier literature] have found this hard to accept
even as a purely gravitational phenomenon, and it is no more palatable when we pass from asymptotically flat to asymptotically AdS spacetimes: recall
that, in doing so, one finds that extremal black
holes cease to be supersymmetric [\cite{kn:clifford}, page 460], so that one no longer has any reason to expect them to be
particularly stable. But it is precisely in the AdS context that
charged black holes are of most physical interest, since these objects are needed
when one applies the AdS/CFT correspondence to strong interaction
or condensed matter physics \cite{kn:chamblin}\cite{kn:gubserreview}\cite{kn:valdivia}\cite{kn:kiritsisreview}\cite{kn:hartnollreview}\cite{kn:hertzogreview}\cite{kn:teaneyreview}.

The problem, as Hartnoll has recently
emphasised [see Footnote 14 of \cite{kn:hartnollreview}], is that the situation seems even stranger on the field theory side of the correspondence than
on the gravitational side. One could imagine that the peculiar physics of black holes might violate the thermodynamic Third Law\footnote{For the [very questionable] current status of a black hole version of the Third Law, see \cite{kn:israel}.}: in fact, not only is it apparently possible to reduce the temperature of the black hole to zero by means of a \emph{finite} sequence of operations, but also the residual entropy can have \emph{any} pre-assigned value. That is, its value is not universal as the Third Law requires. But it is very difficult to believe that all this can happen in any reasonable dual field theory. Turning this around, black holes with well-behaved dual field theories \emph{should} obey the Third Law. In the simplest interpretation, this simply means that zero temperature should not be physically attainable by such black holes.

In the specific case of the holographic description of the strongly coupled Quark-Gluon Plasma [QGP] \cite{kn:valdivia}\cite{kn:teaney}, it is very clear that zero temperature is not attainable in the phase of the field theory that has a black hole dual: there is always a crossover/phase change to the confined phase [with a gravitational dual which is not a black hole], or a phase transition to some other phase [which may not have any gravitational dual at all], as the QGP is cooled: see Figure 1 of \cite{kn:alford}. This strongly suggests that something prevents the dual black hole from reaching extremality.

That there is indeed something peculiar about the extremal limit has been much discussed recently; there are subtle discontinuities \cite{kn:lisa}\cite{kn:garousi} and order-of-limit issues
\cite{kn:mitra} when the limit is taken. These observations may well help to explain why the entropy appears to vanish in some computations. But, as Hartnoll stresses, they are not likely to lead to a resolution of the problem we face here. For extremal black holes
are not alone in causing thermodynamic disquiet: a \emph{near-extremal} black hole has an entropy which is bounded away from zero, but it can have an \emph{arbitrarily}
small temperature as extremality is approached. This is almost as disturbing, on the field theory side of the correspondence, as the situation in the extremal limit itself. Again, in the case of the strongly coupled QGP, the plasma region of the quark matter phase diagram neither reaches nor comes close to the zero-temperature axis.

String theory does contain extremal black holes [with non-zero entropy \cite{kn:vafastrom}]; but the theory itself offers a way to resolve the problems arising from this. The ``weak gravity conjecture" proposed by Arkani-Hamed et al. \cite{kn:vafa} implies that extremal black holes \emph{exist} in string theory but are not \emph{stable}. If this is so, then zero temperature is no longer physically accessible, in practice, by means of a physical variation of the black hole parameters; and so questions as to the properties of extremal black holes become academic. In other words, string theory apparently provides a firm foundation for a genuine thermodynamic Third Law for black holes.

In view of Hartnoll's observation, we propose the following extension of this idea: perhaps black holes, or at least the kinds of
black holes that appear in applications to strong interaction and condensed matter physics, become unstable in one way or another at some point as
extremality is neared, \emph{but before it is reached}. If that is so, then neither zero nor \emph{arbitrarily small} temperatures are physically accessible, by means of finite variations of parameters, for completely [in particular, \emph{non-perturbatively}] stable black holes. In short, stability not only prevents extremality from being attained: it imposes a \emph{non-zero lower bound} on the temperatures of black holes with well-behaved dual field theories.

An interesting way of thinking about this proposal is as follows. Recall [see \cite{kn:clifford}, page 454] that the temperature of an uncharged, spherical, five-dimensional AdS-Schwarzschild black hole is \emph{bounded away from zero}: the black hole decays to an energetically favoured alternative state if one tries to approach zero temperature. This is  interpreted holographically as a deconfinement-confinement transition in the field theory [which in this case is defined on $\bbr$ $\times$ S$^3$]. However, the addition of charge to such a black hole has the effect of
rendering the field theory phase diagram two-dimensional, with the region corresponding to confinement occupying one corner of the diagram [see \cite{kn:clifford}, page 465]. It is now possible, for sufficiently large electric potential at the event horizon [or chemical potential in the field theory], to access the zero-temperature axis while remaining in the deconfined [black hole] phase. What we are suggesting is that black holes dual to physically interesting field theories [defined on \emph{flat} spacetimes] always become unstable when the temperature is sufficiently low. The dual statement is that there is a strip [of non-zero width, which we should be able to compute, or at least adjust to reasonable values], adjacent to the zero-temperature axis in the field theory phase diagram, in which a crossover or phase transition always occurs. \emph{This is in fact a good description of the quark matter phase diagram}, as it is understood phenomenologically \cite{kn:alford}. Indeed, what we are proposing is that it should be possible to construct a quasi-realistic holographic model of the field-theoretic phase diagram.

In order to make this proposal work, we need to do two things: we must identify suitable effects which destabilize black holes that are ``near" to extremality, and we need to define ``nearness" in a precise way.
In this work, we show that we can do these things for the class of charged AdS black holes which are most important in applications: those which are dual to a field theory defined on \emph{flat} spacetime. There are two separate effects acting here. The first is perturbative and relevant at relatively small values of the chemical potential: it is the flat-boundary analogue \cite{kn:horomy}\cite{kn:surya}\cite{kn:gall1}\cite{kn:gall2} of the familiar deconfinement/confinement transition \cite{kn:wittenads} for AdS black holes. The second is non-perturbative and dominates at large values of the chemical potential: it is a purely stringy effect, discovered by Seiberg and Witten \cite{kn:seiberg} [see \cite{kn:porrati} for a clear survey], arising from the possibility of negative free energies for creation of branes. This second effect imposes particularly specific constraints: when it is taken into account, such black holes are stable for almost all values of the charge, but they become unstable when the charge reaches about 96$\%$ of the extremal value. The temperature has a corresponding lower bound, expressible in terms of the black hole mass and certain geometric data at infinity [or in terms of the mass and the minimal possible entropy].

Apart from resolving our problem, considerations like this may put interesting and welcome constraints on the [temperature-normalized] chemical potential of the field theory describing a plasma, $\bar{\mu}$. Recently it has become clear that the celebrated KSS bound \cite{kn:son} can be violated if one goes beyond the two-derivative action in the bulk, and that the introduction of a non-zero $\bar{\mu}$ makes
the violation worse \cite{kn:myers}\cite{kn:cremonini}. Without Seiberg-Witten instability, $\bar{\mu}$ can take arbitrarily large values; with it, we find [using
the normalization of Myers et al. \cite{kn:myers}, and restricting throughout to the two-derivative action] that it is bounded above by about 0.49. While this is only a rough estimate\footnote{While we do not claim that our very elementary discussion is directly related to observations, we note that the value of $\bar{\mu}$ deduced from observations of the unusual states produced at the RHIC [see for example \cite{kn:phobos}] is roughly one-third of our upper bound.}, we can conclude that non-perturbative string effects place very restrictive conditions on $\bar{\mu}$, so that the extent to which the KSS bound can be violated in this way is very effectively contained. This may ultimately point the way to a ``corrected" KSS bound, as we discuss in Section 7.

\addtocounter{section}{1}
\section* {\large{\textsf{2. Physics of ``Flat" Black Holes}}}
In spacetimes satisfying the dominant energy condition ---$\,$ which, for fluid matter, requires that
the energy density equal or exceed the absolute value of the pressure ---$\,$ there are strong restrictions
\cite{kn:greg1}\cite{kn:greg2} on the topology of the event horizon of a black hole. Asymptotically locally anti-de Sitter spacetimes\footnote{For the sake of clarity we shall focus here exclusively on the physically most interesting case, that is, five-dimensional asymptotically locally anti-de Sitter spacetimes dual to four-dimensional field theories.} do
\emph{not} satisfy the dominant energy condition, and this fact does indeed relax these restrictions: now there are genuinely new possibilities for the geometry and topology of the event horizon \cite{kn:lemmo}\cite{kn:mann}\cite{kn:caironggen}\cite{kn:peldan}\cite{kn:danny}. In the case where no matter other than the vacuum energy is present, \emph{all} of the following Anti-de Sitter-Schwarzschild metrics are exact solutions of the Einstein equations\footnote{We stress that, throughout this work, we use the standard Einstein-Hilbert two-derivative action. The consequences of using a four-derivative action are discussed in Section 7.} [with a negative cosmological constant]:
\begin{equation}\label{A}
\m{g(AdSSch_k) = - \Bigg[{r^2\over L^2}\;+\;k\;-\;{16\pi M\over
3\Gamma_kr^2}\Bigg]dt^2\;+\;{dr^2\over {r^2\over L^2}\;+\;k\;-\;{16\pi
M\over 3\Gamma_kr^2}} \;+\; r^2d\Omega_k^2}.
\end{equation}
Here L is the radius of curvature of the asymptotic AdS$_5$,
$\m{d\Omega_k^2}$ is a metric of constant curvature k = $\{- 1, 0, + 1 \}$ on a compact
three-dimensional space C$_{\m{k}}$, and $\Gamma_{\m{k}}$ is the area
of this space. This crucial quantity enters here because of the normalization of the ADM Hamiltonian
at infinity: if this Hamiltonian is to vanish at zero mass [and charge] then \cite{kn:peldan} one needs this factor at each appearance of the mass and charge\footnote{Throughout
this work, M denotes the ADM mass, and Q the ADM charge; for simplicity we shall take it that Q is positive.}.

The cases with k $\neq + 1$ are sometimes called ``local" versions of the k = 1
spacetime, presumably by analogy with the forms taken by the de Sitter metric on submanifolds of de Sitter spacetime which can be foliated by flat or
hyperbolic spatial sections [see for example \cite{kn:holography}], or perhaps because Witten \cite{kn:confined} first obtained the k = 0 metric [in this context] by taking the large-mass limit of the k = 1 spacetime. This terminology is however
extremely misleading: the k $\neq + 1$ metrics are \emph{exact} solutions at \emph{any} value of the mass, large or small; and their
global structure is completely different from that of any black hole with a locally spherical event horizon. In particular, the conformal
boundary has the topological and conformal structure of $\bbr \,\times $ C$_{\m{k}}$, which differs radically from one value of k to another.

In applications of the AdS/CFT correspondence to strong interaction and condensed matter physics, one is almost exclusively interested in defining the field theory on a spacetime which is locally indistinguishable from Minkowski spacetime. It follows that the black holes which appear in these
applications are those with metrics given by equation (\ref{A}) with k = 0. In view of the comments we have just made, we must
not assume that these black holes behave in precisely the same way as their more familiar counterparts
with locally spherical event horizons: they are not ``local versions" of the latter. Let us therefore be more precise
about the nature of the k = 0 case.

The space C$_0$ is not uniquely defined \cite{kn:conway}, but, for the sake of simplicity, let us assume that C$_0$ is an exactly flat torus T$^3$, parametrised by three angles. We shall assume that this torus is cubic: that is, all three angles have periodicity 2$\pi$K, where K is a dimensionless parameter. Then the quantity $\Gamma_0$ in equation (\ref{A}) is given by
\begin{equation}\label{B}
\m{\Gamma_0\;=\;8\,\pi^3\,K^3},
\end{equation}
and the area of any surface defined by r = constant is equal to this quantity multiplied by r$^3$.

Thus the metric in which we are interested is
\begin{equation}\label{C}
\m{g(AdSSch_0) = - \Bigg[{r^2\over L^2}\;-\;{2M\over
3\pi^2K^3r^2}\Bigg]dt^2\;+\;{dr^2\over {r^2\over L^2}\;-\;{2M\over
3\pi^2K^3r^2}} \;+\; r^2d\Omega_0^2}.
\end{equation}

This metric has a Euclidean version,
in which the complexified version of t parametrises a [fourth]
circle; thus the Euclidean metric, g(EAdSSch$_0$), is a metric
on a manifold which is radially foliated by copies of the four-torus T$^4$. One can think of t/L as an angular
coordinate on the fourth circle; the periodicity 2$\pi$P of this angular coordinate must be chosen so
that the Euclidean metric is
not singular at r$_{\m{eh}}$, the value of r at the event horizon. The conformal boundary now has the
structure of a [conformal] torus, which however need not be cubic, since in general P will not be equal to K. The field theory is defined on this space, which is just [the
Euclidean version of] ordinary flat spacetime \cite{kn:confined} with formal periodic boundary conditions . [The Lorentzian topology at infinity is of course $\bbr \;\times$ T$^3$; the structure of infinity is necessarily related to that of the event horizon by topological censorship \cite{kn:topcensor}.]

A non-zero chemical potential, $\mu$, is needed if the field theory is to have a continuously variable temperature;
as is well known [see for example \cite{kn:ge1}\cite{kn:hartnollreview}] this corresponds to putting a charge on the black hole
on the gravitational side of the AdS/CFT correspondence. Let us therefore discuss the charged version of the above ``flat" black hole, and then show how to express the chemical potential in terms of black hole parameters.

If the ADM charge is Q, then the AdS-Reissner-Nordstr$\ddot{\m{o}}$m metric with flat event horizon is
\begin{equation}\label{D}
\m{g(AdSRN_0) = - \Bigg[{r^2\over L^2}\;-\;{2M\over
3\pi^2K^3r^2}\;+\;{Q^2\over 48\pi^5 K^6 r^4}\Bigg]dt^2\;+\;{dr^2\over {r^2\over L^2}\;-\;{2M\over
3\pi^2K^3r^2}\;+\;{Q^2\over 48\pi^5 K^6 r^4}} \;+\; r^2d\Omega_0^2};
\end{equation}
the power of K in the new terms is required by the fact that Q appears as a square.

The value of r at the event horizon, assuming that one exists, is obtained by solving the equation
\begin{equation}\label{E}
\m{{r_{eh}^6\over L^2}\;-\;{2Mr_{eh}^2\over
3\pi^2K^3}\;+\;{Q^2\over 48\pi^5 K^6} \;=\;0.}
\end{equation}
The value of r$_{\m{eh}}$ is mainly of interest because of its relation to the black hole entropy. As was recently demonstrated in detail in \cite{kn:lemos2},
the standard relation between the entropy of a black hole and its horizon area continues to hold for these black holes: that is, the entropy is one
quarter of the horizon area, so that
\begin{equation}\label{F}
\m{S \;=\;2\pi^3K^3r_{eh}^3.}
\end{equation}
A straightforward calculation now allows us to eliminate r$_{\m{eh}}$ between (\ref{E}) and (\ref{F}). This allows us to fix the entropy in terms of the values of the ADM and geometric parameters: the entropy satisfies
\begin{equation}\label{G}
\m{\pi Q^2L^2 \;=\;2^{1/3} \times 16 \pi^2 M L^2 K S^{2/3} \;-\; 12S^2.}
\end{equation}
``Approaching extremality" is most simply implemented as a process whereby one fixes all other ADM and geometric parameters and imagines increasing the charge. From (\ref{G}), we see that Q$^2$ can be regarded as a function of S, and vice versa. Notice that this apparently maps, in general, two values of S to a given value of Q$^2$, but this is due to the fact that (\ref{E}) yields
two values for r$_{\m{eh}}$ when the other parameters are fixed: in the non-extremal case there are two Reissner-Nordstr$\ddot{\m{o}}$m horizons, as usual.
The smaller value corresponds to a Cauchy horizon; only the larger value defines an event horizon. Therefore, in computing the entropy we should only
take the larger of the two candidates for r$_{\m{eh}}$, and from (\ref{F})
we see that this implies that we accept only the larger value of S. That is, only the right-hand [decreasing] branch of the graph of Q against S is physical. Figure 1 portrays Q$^2$ as a function of S in the physical domain.
\begin{figure}[!h]
\centering
\includegraphics[width=1.2\textwidth]{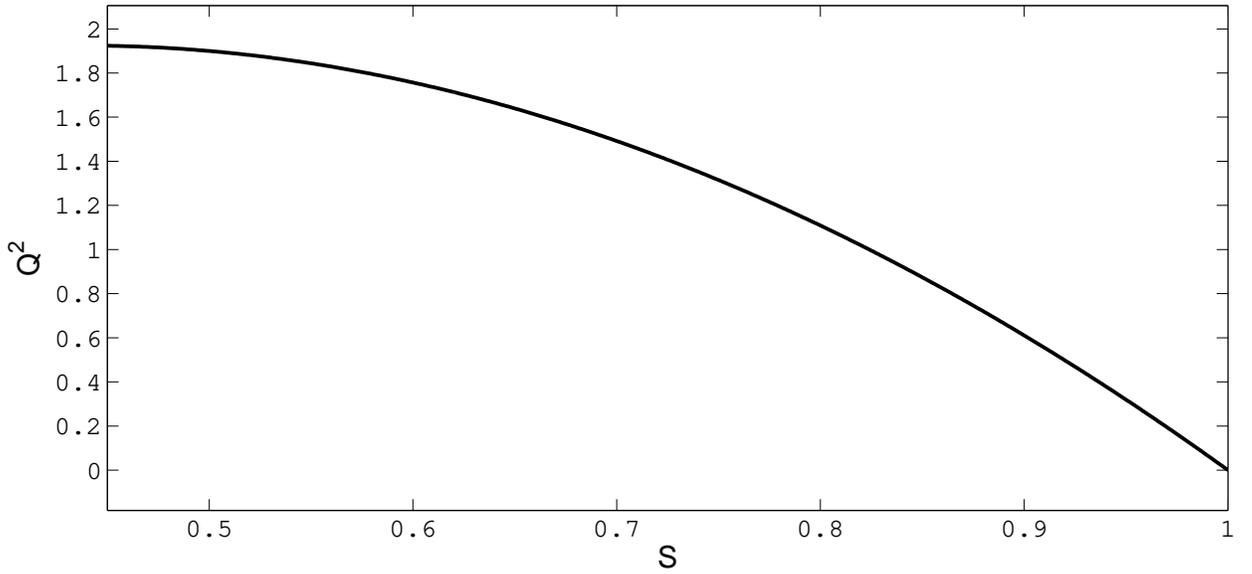}
\caption{Squared Charge as a Function of Entropy [Typical Parameter Values]}
\end{figure}
The key point here is that the charge reaches its maximum allowed value before the entropy is able to reach zero. That is, the vertical axis in Figure 1 does not correspond to S = 0. There is a lower bound on the possible values of the entropy: the entropy cannot be smaller than its value when the black hole is extremal:
\begin{equation}\label{H}
\m{S\;\geq\;S_{E}\;=\;\Bigg[{4 \times 2^{1/3}\over 9}\,\times \,\pi^2 M L^2 K\Bigg]^{3/4}\;\approx \;0.647322\times \Big[\pi^2 M L^2 K\Big]^{3/4}.}
\end{equation}
There is also an upper bound on the entropy, given by its value for an uncharged ``flat" black hole: the addition of charge always reduces the entropy [all other parameters being fixed]. The lower bound on the entropy implies that there is an upper bound on the charge, the charge of the extremal black hole:
\begin{equation}\label{I}
\m{Q^2\;\leq\;Q^2_{E}\;=\;{64\over 9}\,\sqrt{2}\pi^2 M^{3/2} L K^{3/2}.}
\end{equation}
Because it is easier to express Q$^2$ in terms of S rather than the reverse, we shall often find it convenient to use S as our independent parameter, but the reader can always think in terms of varying Q, simply by bearing Figure 1 in mind.

The bounds on the charge and entropy are saturated when the two possible values of r$_{\m{eh}}$ [for
given values of M, Q, L, and K] actually coincide: that is, they are saturated only by an extremal black hole. Beyond that point, r$_{\m{eh}}$ is not well-defined. Thus, the fact that Q$^2$ is bounded above, and that S is bounded below, is a reflection of the assumed existence of an event horizon. Classically, one would describe this situation in terms of ``cosmic censorship". [See \cite{kn:hod}\cite{kn:lemos1}\cite{kn:saa}, but also \cite{kn:gubser}.]

The temperature of the black hole is computed by evaluating 2$\pi$P, the periodicity of the complexified time coordinate, and the result may be written [with the aid of equations (\ref{E}),(\ref{F})] as
\begin{equation}\label{J}
\m{T_{}\;=\;{1\over  2^{1/3}\pi}\Bigg[{S^{1/3}\over \pi KL^2}\;-\;{Q^2 \over 24K S^{5/3}}\Bigg].}
\end{equation}
Now Q can be eliminated using equation (\ref{G}), and after a lengthy simplification we obtain
\begin{equation}\label{K}
\m{T_{}\;=\;{3S^{1/3}\over 2^{1/3}\times 2\pi^2 KL^2}\;-\;{2M\over 3S}.}
\end{equation}
This function is portrayed in Figure 2. The temperature always increases with the entropy. For large S [small Q] the temperature varies with the cube root of the entropy, just as it does for uncharged spherical five-dimensional AdS black holes [see \cite{kn:clifford}, page 454]; the system does not ``care" about the spatial geometry when the entropy is large, as is intuitively reasonable since a large sphere resembles a large torus locally\footnote{In both cases, ``large" entropy is associated with large mass: recall that the entropy of the uncharged case, fixed by the mass, sets an upper bound to the entropy for charged ``flat" black holes.}. But for small S [large Q] the temperature vanishes before the entropy does so, reaching zero when the entropy is given by its extremal value S$_{\m{E}}$ [equation (\ref{H})]. This is our problem: Figure 2 looks reasonable from the field theory point of view when S is large, but not when it is small.
\begin{figure}[!h]
\centering
\includegraphics[width=1.2\textwidth]{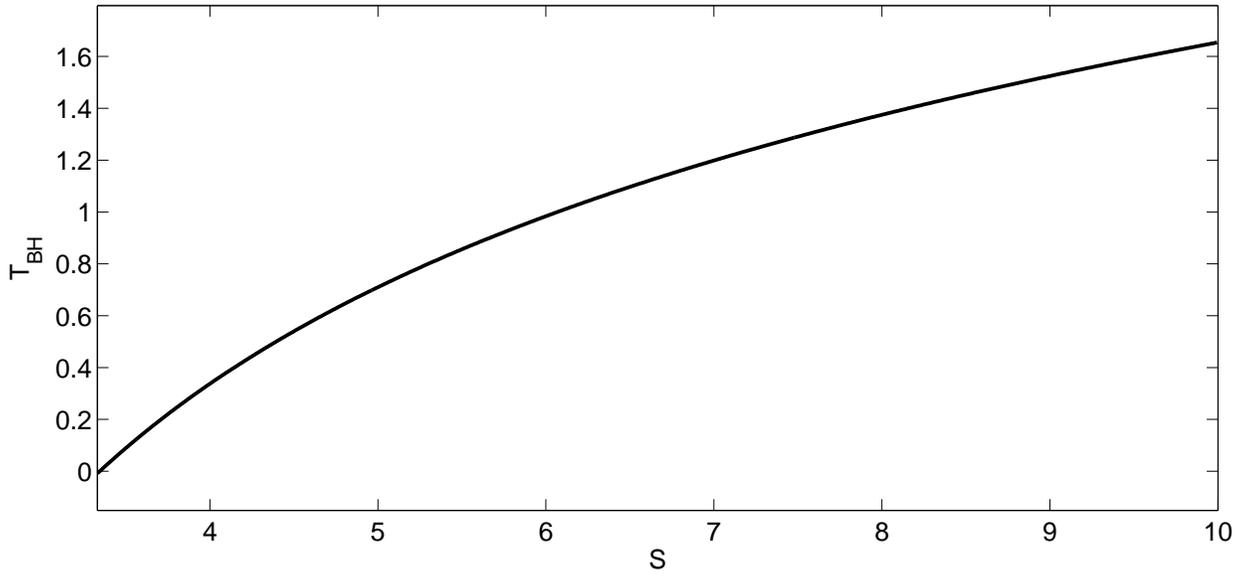}
\caption{Black Hole Temperature as a Function of Entropy [Typical Parameter Values]}
\end{figure}

We are now in a position to compute the chemical potential. The bulk electromagnetic potential one-form, expressed in terms of a gauge
such that the connection is not singular [see \cite{kn:clifford}, page 416], is
\begin{equation}\label{L}
\m{A\;=\;{Q \over 16\pi^3K^3}\Bigg[{1\over r^2}\;-\;{1\over r_{eh}^2}\Bigg]\,dt.}
\end{equation}
The chemical potential $\mu$ on the field theory side of the AdS/CFT correspondence [see \cite{kn:hartnollreview}] is
proportional to the background electric potential corresponding to a global U(1) symmetry. The [negative] constant of proportionality must have units of inverse length if the chemical potential is to have the correct units; here we adopt the procedure followed in \cite{kn:myers}. The chemical potential is therefore
given by the product of this constant with the magnitude of the asymptotic value of the coefficient of dt in equation (\ref{L}):
\begin{equation}\label{M}
\m{\mu\;=\;{Q \over 16\pi^4K^3L r_{eh}^2}\;=\;{2^{2/3}Q \over 16\pi^2 K L S^{2/3}}.}
\end{equation}
If we think of S as a solution of equation (\ref{G}), then we see that the chemical potential is determined not just by the charge but also by the bulk black hole's mass and its spatial periodicity parameter K. It follows that any restrictions on these quantities impose restrictions on the chemical potential. For example, if we fix M and K and allow Q [and therefore S] to vary, the assumed existence of a well-defined entropy on the AdS side of the duality imposes restrictions on Q and S [inequalities (\ref{H}) and (\ref{I})]. Since S is a decreasing function of Q under these circumstances [Figure 1], we see that $\mu$ always increases with Q: it is zero when Q vanishes, and it is bounded above by its value at extremality:
\begin{equation}\label{N}
\m{\mu\;\leq\;\mu_{E}\;=\;{1\over 2^{5/4} \pi^2}\,\Bigg[{M\over K^3L^6}\Bigg]^{1/4}\;\approx \;0.042600\times \Bigg[{M\over K^3L^6}\Bigg]^{1/4}.}
\end{equation}
Any further constraints on Q will of course amend this upper bound.

Following \cite{kn:myers} once more, we define the chemical potential normalized by the temperature,
\begin{equation}\label{NN}
\m{\bar{\mu}\;=\;\mu/T.}
\end{equation}
Because $\mu$ is an increasing, and T a decreasing function of Q, we see that $\bar{\mu}$ is an increasing
function of Q [assumed positive for convenience]; unlike $\mu$, however, it is \emph{not} classically bounded above. This is another way of stating our basic concern, the possibility of having black holes
with zero temperature; a solution of this problem will entail an upper bound on $\bar{\mu}$. We record this crucial point as follows: the range of $\bar{\mu}$ is given classically by
\begin{equation}\label{NNN}
\m{0\;\leq \;\bar{\mu}\;<\;\infty.}
\end{equation}
We shall see later that string theory drastically modifies this statement.

\addtocounter{section}{1}
\section* {\large{\textsf{3. No Zero Temperatures at Low $\mu$ : Transition to Confinement}}}

Let us focus now on a particular application, the use of these black holes to describe the quark-gluon plasma \cite{kn:valdivia}. In phenomenological studies \cite{kn:alford}, the phase diagram is a graph of the temperature T against the chemical potential $\mu$. Holographically, these two parameters are, as we have seen, just complicated functions of the black hole ADM parameters M and Q; see equations (\ref{J}) and (\ref{M}). Phase changes and crossovers in the ($\mu$, T) plane are a result of changes in the behaviour of the the black hole when M and Q are allowed to vary. These changes are due in part to the thermodynamics of the black hole, to which we now turn.

The thermodynamic behaviour of these so-called ``flat" black holes is not simple. One should \emph{not} use the zero-mass limit as the background for regularizing the energy, but rather the Horowitz-Myers ``AdS soliton" \cite{kn:horomy}; for it was conjectured, and subsequently proved, that this is the configuration of least energy with these boundary conditions \cite{kn:surya}\cite{kn:gall1}\cite{kn:gall2}. The various phase transitions depend not just on the ADM parameters of the black hole but also on the fixed geometric parameters K and L. To explain this, we write out the d$\Omega_0^2$ term in equation (\ref{D}) in full, taking the coordinates on the three-torus to be $\phi$, $\theta_1$, $\theta_2$ [that is, we arbitrarily single out one of the toral directions], so that the black hole metric is
\begin{eqnarray}\label{NNNN}
\m{g(AdSRN_0)} &=& -\, \m{\Bigg[{r^2\over L^2}\;-\;{2M\over
3\pi^2K^3r^2}\;+\;{Q^2\over 48\pi^5 K^6 r^4}\Bigg]dt^2\;}
\nonumber \\
& & + \m{\;{dr^2\over {r^2\over L^2}\;-\;{2M\over
3\pi^2K^3r^2}\;+\;{Q^2\over 48\pi^5 K^6 r^4}} \;+\; r^2\,\Big[d\phi^2\;+\;d\theta_1^2\;+d\theta_2^2\Big]}.
\end{eqnarray}

The AdS-Reissner-Nordstr$\ddot{\m{o}}$m soliton metric is then obtained by means of a straightforward double analytic continuation of the black hole metric: complexifying t and $\phi$ and exchanging labels in a natural way we obtain\footnote{The basic references \cite{kn:horomy}\cite{kn:surya}\cite{kn:gall1}\cite{kn:gall2} deal only with the uncharged case; we are supposing that the basic ideas still hold here.}
\begin{eqnarray}\label{NNNNN}
\m{g(AdSRNSol)} &=& -\, \m{{r^2\over L^2}\,dt^2\;
+\;{dr^2\over {r^2\over L^2}\;-\;{A\over
r^2}\;+\;{B\over r^4}}}
\nonumber \\
& & \m{\;+\; \Bigg[{r^2\over L^2}\;-\;{A\over
r^2}\;+\;{B\over r^4}\Bigg]\,L^2\,d\phi^2\;+\;r^2\,\Big[d\theta_1^2\;+d\theta_2^2\Big]}.
\end{eqnarray}
As was emphasised in \cite{kn:surya}, there is no need to insist that the parameters here should coincide with those of the black hole. However, the angular coordinates are well-defined at conformal infinity, so their periodicities must agree with those of the hole in order that the two geometries should match there. As in the black hole case, the need to avoid a conical singularity imposes a relationship between periodicities and parameters. In the black hole case, this fixes the temperature in terms of the black hole parameters; here we have to apply this idea to $\phi$, which has periodicity
2$\pi$K. This gives us one relation between A, B, and K. [For our purposes it is best to think of B as being fixed at some small value; then A is determined by K. This allows the soliton to avoid the form of instability discussed in the next section.] Finally,
the Euclidean ``time" coordinate of the soliton has to have a periodicity which matches that of the Euclidean black hole ``time" coordinate, namely, 2$\pi$P.

As usual, the Euclidean action of these black holes is divergent and has to be regularized by means of a comparison with a reference spacetime, which in this case we take to be a Horowitz-Myers soliton. One can show, following \cite{kn:surya}, that the regularized black hole action is given in this [five-dimensional] case by
\begin{equation}\label{NNNNNN}
\m{I \;=\;\alpha\,PK^3\,\Big[K^{-\,4}\;-\;P^{-\,4}\Big];}
\end{equation}
here $\alpha$ is a positive number [which depends on L], whose precise value does not concern us. If we had insisted that the soliton parameters coincide with those of the black hole, then K and P would have been given by identical formulae, and so this action would have vanished and there would have been no possibility of any phase transitions. But if we allow other values for A and B in equation (\ref{NNNNN}), then it becomes possible for K to differ from P, and phase transitions can and do occur. [The action depends on A and B, but only through K; alternatively, one can, as above, think of K as the basic variable, fixing B appropriately and regarding A as a function of K.] The phase transitions are controlled by the relative sizes of K and P: in other words, they are determined by the precise shape of the four-torus at [Euclidean] conformal infinity, by the extent to which it deviates from being cubic.

Writing P as 1/(2$\pi$TL), where T is the temperature of the black hole, we have
\begin{equation}\label{NNNNNNN}
\m{I \;=\;\alpha\,PK^3\,\Big[K^{-\,4}\;-\;(2\pi TL)^4\Big].}
\end{equation}
Assuming that the free energy of the soliton is zero, we see now that the black hole is energetically favoured only if its temperature is not too low compared with 1/(2$\pi$KL). This means that if K is very large, then a very cold black hole can be stable; conversely, if K is very small, then even a very hot black hole can be unstable, as was pointed out in \cite{kn:surya}. The dual interpretation is that the phase of the field theory is determined by the size of the temperature relative not to 1/L but rather to 1/KL, where K is a purely geometric parameter which cannot be varied by means of any local physical process.

While a very cold ``flat" black hole can be stable in some special cases, the fact remains that \emph{zero} temperature is impossible for these black holes. Once K is fixed, every ``flat" black hole with a temperature below 1/(2$\pi$KL) will be unstable: it will be replaced by the appropriate soliton, which then has lower free energy. So the temperature of the hole satisfies
\begin{equation}\label{NNNNNNNN}
\m{T\; \geq \; {1 \over 2\pi KL}.}
\end{equation}
This is of course exactly what we requested in Section 1: the hole becomes unstable as it is cooled, and this happens even before extremality is reached. The dual statement is that the plasma phase of the field theory cannot be arbitrarily cold.

Generally speaking, ``flat" black holes are easier to interpret, from a holographic point of view, than their spherical counterparts. For example: all ``flat" black holes have \emph{positive specific heat} under all circumstances, as discussed in \cite{kn:surya} in the uncharged case ---$\,$ the black hole is \emph{eternal}\footnote{Of course it will only be literally eternal if it can reach equilibrium with its own Hawking radiation before its temperature falls too low.} \cite{kn:juan}. The positivity of the specific heat is very reasonable from the dual point of view, since one certainly expects the specific heat of the field theory to be positive \cite{kn:confined}.

The case at hand provides another example. A spherical AdS-Reissner-Nordstr$\ddot{\m{o}}$m black hole has a critical value of the charge [\cite{kn:clifford}, page 461] above which zero temperature \emph{is} possible. In the phase diagram [\cite{kn:clifford}, page 465] this means that, in the spherical case, the confined field theory corresponds to one small \emph{corner} of the diagram. In the ``flat" case, by sharp contrast, it is represented by an infinite \emph{strip}, of width 1/(2$\pi$KL), adjacent to the T = 0 axis.

That is a very considerable improvement over the spherical case, since the ``schematic" quark matter phase diagram given in \cite{kn:alford} does exhibit such a strip, forbidding zero temperature to the plasma phase; but it is still unacceptable. For it predicts that the plasma phase of the field theory makes a phase transition to a confined state [corresponding to the soliton], as the temperature is lowered, at \emph{all} values of the chemical potential, no matter how large. A glance at the quark matter phase diagram, even if we accept that it is only qualitatively valid, shows that this is incorrect. At high values of the chemical potential, there is a phase transition, but \emph{not} to a confined state; these new states, such as the ``colour-flavour locked" state, apparently have no gravitational dual description. A convincing holographic account of the phase diagram therefore cannot rely exclusively on the effect we have been discussing in this section. We need another effect, one that dominates at high chemical potential. We now turn to this.

\addtocounter{section}{1}
\section* {\large{\textsf{4. No Zero Temperatures at High $\mu$ : Stringy Instabilities}}}
It has been understood from the beginning \cite{kn:wittenads} that the AdS/CFT correspondence should be valid for
more general asymptotically locally AdS spacetimes, not just for AdS itself. In physical language this means the following. Suppose that we introduce matter into AdS, and allow it to deform the spacetime geometry in accordance with either classical gravity or some stringy modification of it. If the deformation is sufficiently mild, the spacetime will continue to have a conformal compactification with a Lorentzian conformal structure at infinity. Even though the geometry of conformal infinity may change quite substantially, it is reasonable in this case to assume that there continues to be a duality between a field theory at infinity and the gravitational theory in the bulk.

However, this does not mean that every deformation of AdS with a Lorentzian conformal boundary is physically acceptable. Extending earlier work in more
restrictive circumstances \cite{kn:frag}, Seiberg and Witten \cite{kn:seiberg}, and Witten and Yau \cite{kn:wittenyau}, showed that if the boundary is deformed to such an extent that the scalar curvature of its Euclidean version becomes \emph{negative}, then the entire system becomes unstable. [The scalar curvature at infinity is positive for AdS itself.] They did this by computing the action of a ``large" BPS brane. This turns out to become, and remain, negative at some finite distance if the scalar curvature at infinity is negative, giving rise to an infinite reservoir of negative free energy. This means that there is a pair-production instability for branes. Note here that extended objects like branes are capable of detecting at least some aspects of the global structure of spacetime, so one must take care to use the correct global form of the spacetime, and not to be misled by special, local choices of coordinates [such as those which suggest that the scalar curvature of the Euclidean AdS conformal boundary is \emph{zero}].

It is important to understand that the analysis of Witten and Yau is extremely general: though they focus on the case in which the bulk is an Einstein manifold, they note that their methods, including their computation of the brane action, remain valid for any asymptotically AdS spacetime. For example,
Maldacena and Maoz \cite{kn:maoz} studied Seiberg-Witten instability in the presence of a gauge field, using the Witten-Yau expression for the BPS brane action; here the bulk is asymptotically locally AdS but not an Einstein manifold\footnote{Witten and Yau, and also Maldacena and Maoz, were concerned with the issue as to whether the boundary could be topologically disconnected in the AdS/CFT correspondence; this question was clarified subsequently by Cai and Galloway \cite{kn:galloway}; see also \cite{kn:malda}. This particular issue does not arise here.}. This is of course precisely the case of interest to us here, and we shall proceed in the same way.

This ``Seiberg-Witten instability" certainly arises at large values of a radial coordinate if the scalar curvature at infinity is negative, but not if the scalar curvature at infinity is positive. Thus for example AdS black strings dual to a gauge theory on S$^2 \times$ S$^1 \times \bbr$
\cite{kn:GTH} are stable. There remains however a third possibility: if the scalar curvature at infinity is strictly \emph{zero}, then the brane action \emph{might} become negative at large distances,
depending on higher-order terms in the expansion considered by Seiberg and Witten. Of course, asymptotically locally AdS black holes with flat event horizons have zero scalar curvature at infinity, so they lie precisely in this extremely delicate ``borderline" region. We should expect that some of these black holes are stable, while others are pushed over the borderline when some parameter is continuously adjusted. As we shall see, that is precisely what happens.

It was shown in \cite{kn:conspiracy} that uncharged AdS black holes with flat event horizons are in fact \emph{stable} in the Seiberg-Witten sense. We wish to extend that discussion to the charged case. To investigate this, we switch to the Euclidean picture, and assume as usual that the temporal period 2$\pi$P
has been chosen so that the Euclidean version of the charged black hole spacetime with metric (\ref{D}) is
non-singular. We can consider a BPS 3-brane of tension $\Theta$ wrapping one of the r = constant sections of that space; the action is then, following \cite{kn:seiberg}\cite{kn:wittenyau}\cite{kn:porrati},
\begin{equation}\label{P}
\m{\$(r ; \Theta, L, M, Q, K) \;=\;16\pi^4 \Theta P L K^3 \Bigg\{
r^3\Bigg[\,{r^2\over L^2}\;-\;{2 M\over
3\pi^2 K^3 r^2}\;+\;{Q^2\over 48\pi^5 K^6 r^4}\,\Bigg]^{1/2}\;-\;{r^4\,-\,r_{eh}^4\over L}\Bigg\}.}
\end{equation}
Here the notation means that the action is a function of r with parameters $\Theta$, L, M, Q, and K; note that r$_{\m{eh}}$ and P are fixed by these parameters. [r$_{\m{eh}}$ is found by solving (\ref{E}); P is proportional to the reciprocal of the temperature, given by equation (\ref{K}); it depends on S, which however is fixed by L,M,Q, and K.]

This expression simplifies to
\begin{equation}\label{Q}
\m{\$(r ; \Theta, L, M, Q, K) \;=\;16\pi^4\Theta P L^2 K^3\Bigg\{{{Q^2 \over 48\pi^5 K^6 r^2}\;-\;{2M \over 3\pi^2 K^3}
\over
1\;+\;\Bigg[\,1\;-\;{2 ML^2\over
3\pi^2 K^3r^4}\;+\;{Q^2L^2\over 48\pi^5 K^6 r^6}\,\Bigg]^{1/2}}\;+\;{r_{eh}^4\over L^2}\Bigg\}}.
\end{equation}
This function vanishes at r = r$_{\m{eh}}$, both the area and the volume of the brane being zero there; it then becomes positive, but is asymptotic at large r to the value
\begin{equation}\label{R}
\m{\$(\infty ; \Theta, L, M, Q, K) \;=\;16\pi^4\Theta P L^2 K^3\Bigg\{- {M \over 3\pi^2 K^3}
\;+\;{r_{eh}^4\over L^2}\Bigg\}}.
\end{equation}

We remind the reader that this asymptotic action is computed using the Einstein-Hilbert gravitational action [which leads to the metric (\ref{D})]. In other words, following Witten, Yau \cite{kn:wittenyau}, Maldacena, and Maoz \cite{kn:maoz}, we ignore all couplings of the probe brane and all stringy corrections to the gravitational action. Our results are therefore approximate. Since we are trying to exclude an entire \emph{range} of values for Q, and not just a single [extremal] value, we believe that this will not affect our main conclusions: it will only give rise to a correction to the \emph{width} of the range of excluded charges. In short, the qualitative picture should be valid\footnote{To establish this more firmly, one should re-compute the brane action in the context where the gauge field is explicitly obtained from a Kaluza-Klein reduction of one of the dimensions we have ignored here, and with a coupling of the brane to the RR flux. We intend to return to this issue elsewhere.}. For the consequences of going beyond the Einstein-Hilbert action, see Section 7 below.

It is easy to show that the asymptotic action in (\ref{R}) is positive if Q = 0, and this is why the uncharged black hole is stable. The addition
of electric charge has, however, the effect of \emph{reducing} r$_{\m{eh}}$: the values of r at the two horizons approach each other as extremality is neared. Thus there is a danger that the negative term in the braces could outweigh the positive term in the charged case. This is precisely what happens, and it happens \emph{before} extremality is reached. We are now ready to be precise about this.

\addtocounter{section}{1}
\section* {\large{\textsf{5. The Bounds on Temperature and Normalized Chemical Potential}}}
We can assess the relevant values of the asymptotic brane action in an elementary and explicit way as follows. Let x = r$^2$, and x$_{\m{eh}}$ = r$^2_{\m{eh}}$.
Then [see equation (\ref{E})] x$_{\m{eh}}$ is [when the two horizon radii are distinct] the larger of the two positive roots of the cubic
\begin{equation}\label{S}
\m{G(x)\;=\;x^3\;-\;{2ML^2x\over
3\pi^2K^3}\;+\;{Q^2L^2\over 48\pi^5 K^6}.}
\end{equation}
This cubic has its minimum at
\begin{equation}\label{T}
\m{x_{min}\;=\;\Bigg[{2ML^2\over 9\pi^2 K^3}\Bigg]^{1/2}}.
\end{equation}
The condition for an event horizon to exist is that G(x$_{\m{min}}$) should not be positive. A straightforward calculation shows that this demands that the following dimensionless combination of the parameters should satisfy
\begin{equation}\label{U}
\m{{Q^2\over L(KM)^{3/2}}\;\leq \; {64\sqrt{2}\pi^2 \over 9}\;\approx\;99.2550};
\end{equation}
this is of course another form of (\ref{I}), and the inequality is again saturated only at extremality. We continue to denote the extremal value of Q$^2$ by Q$^2_{\m{E}}$.

As usual, we shall vary Q while keeping M, L, and K fixed\footnote{We assume for simplicity that, as this happens, the system remains in the plasma phase [that is, that the black hole does not dissolve into a soliton]. See Section 6.}. Note that, as we do this, x$_{\m{min}}$ \emph{remains fixed}. Increasing Q simply has the effect of lifting the cubic higher, without changing the location of its minimum point. This draws the two positive roots together: in other words, r$_{\m{eh}}$ becomes steadily smaller as Q increases [and hence so does S, in agreement with our earlier discussions]. The two positive roots of the cubic G(x) will coalesce at x$_{\m{min}}$ in the extremal limit, which means that x$_{\m{min}}$ is the square of r$_{\m{eh}}$ in the extremal case; so we can try to evaluate the brane action at infinity for the extremal black hole simply by letting r$_{\m{eh}}$ tend to the square root of x$_{\m{min}}$ in equation (\ref{R}). The result is divergent, because the periodicity P diverges in the extremal limit; but the ratio $\m{\$(\infty ; \Theta, L, M, Q, K)}$/P has a finite limit as extremality is approached, given by
\begin{equation}\label{V}
\m{\lim_{Q
\rightarrow Q_{E}}\$(\infty ; \Theta, L, M, Q, K)/P \;=\;-\,{16\over 9}\pi^2\Theta L^2 M}.
\end{equation}
This is of course negative\footnote{Notice that the result depends on M but not on K; the result is independent of the periodicities of the coordinates.}; we can interpret this as meaning that when the temperature is extremely low, so that P is very large, the brane action at infinity is an arbitrarily large \emph{negative} number. In other words, \emph{near-extremal black holes dual to strongly coupled field theories on flat spacetime are unstable} in the Seiberg-Witten sense.

We can be more precise about ``nearness", as follows. Define a critical [``near-extremal"] value of Q$^2$ by
\begin{equation}\label{W}
\m{{Q_{NE}^2\over L(KM)^{3/2}} \;=\;{16\pi^2 \over \sqrt{3}}\;\approx\;91.1715};
\end{equation}
note the crucial fact that Q$^2_{\m{NE}}$ is smaller than Q$^2_{\m{E}}$, so that (\ref{U}) is satisfied:
in fact,
\begin{equation}\label{X}
\m{Q^2_{NE}/Q^2_{E}\;\approx\;0.918559},
\end{equation}
so we are below extremality in this case, though not far below it: the charge is at about 96$\%$ of its extremal value. We now claim that, for this near-extremal combination of parameters, x$_{\m{eh}}$ is given by
\begin{equation}\label{Y}
\m{x_{eh}(NE)\;=\;\Bigg[{ML^2\over 3\pi^2 K^3}\Bigg]^{1/2}}.
\end{equation}
To see this, one simply has to verify directly, using the definition of Q$^2_{\m{NE}}$, that G(x) vanishes at this value, and to check that x$\m{_{eh}(NE)\,>\,x_{min}}$ [so that we are indeed dealing with the \emph{larger} of the two positive roots of the cubic]. This is elementary. But now if we substitute the corresponding value of r$_{\m{eh}}$ into equation (\ref{R}), then we find that the asymptotic brane action \emph{vanishes}. This is the sense in which Q$^2_{\m{NE}}$ is a critical value for the squared charge: the system is in this case on the brink of becoming unstable. This means that if [again, fixing M, L, and K] we increase Q beyond this critical value, then [see equations (\ref{S}) and (\ref{T})] the graph of the cubic rises up, reducing r$_{\m{eh}}$ still further. But then (\ref{R}) implies that the brane action at infinity becomes negative: in short, the system is unstable for any value of Q$^2$ strictly greater than Q$^2_{\m{NE}}$. This gives a precise definition of ``near-extremal".

Figure 3 shows the brane action for a typical choice of parameters, such that Q is approximately 98.8$\%$ of its extremal value; that is, we are well inside the ``near-extremal" regime, but \emph{not} at extremality. The fact that the action eventually becomes negative is clear.
\begin{figure}[!h]
\centering
\includegraphics[width=1.2\textwidth]{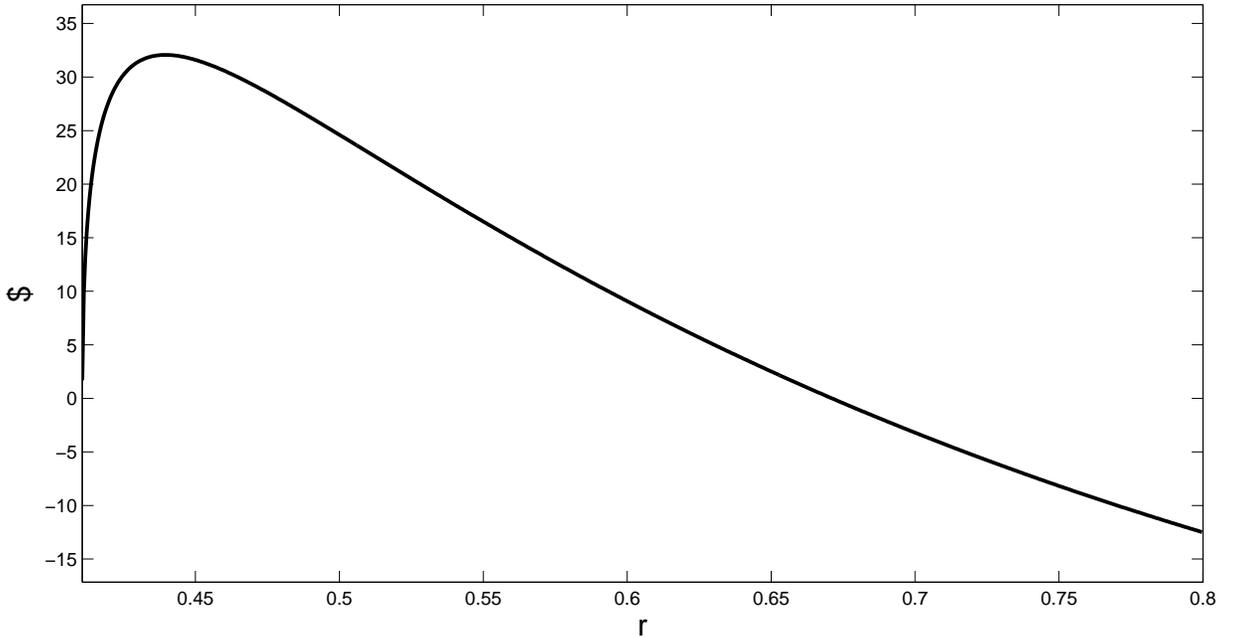}
\caption{Brane Action, Typical Parameters, With Q at 98.8$\%$ of Extremality}
\end{figure}

To summarize: for most values of the charge, these black holes are stable in the Seiberg-Witten sense. But if the charge is increased beyond approximately 96$\%$ of the extremal value, then they cease to be stable. For simplicity, we have stressed the role of the charge; but similar conclusions hold [see equation (\ref{W})] if either the mass or the spatial periodicity is too small.

Since Q and S are related when the other parameters are fixed [Figure 1], it follows that a more restrictive upper bound on Q will impose a lower bound on S which is more restrictive than the one given by the inequality in (\ref{H}), above. Let S$_{\m{NE}}$ be the value of the entropy which corresponds to the ``near-extremal" value of the charge; then we have, from equation (\ref{Y}),
\begin{equation}\label{Z}
\m{S_{NE}\;=\;2 \pi^3 K^3 [x_{eh}(NE)]^{3/2}\;=\;{2\over 3^{3/4}}\times \Big[\pi^2 M L^2 K\Big]^{3/4}.}
\end{equation}
Thus we have
\begin{equation}\label{ALPHA}
\m{S\;\geq\;S_{NE}\;\approx\; 0.877383\times \Big[\pi^2 M L^2 K\Big]^{3/4}.}
\end{equation}
This is indeed significantly more restrictive than the inequality in (\ref{H}).

The temperature corresponding to the ``near-extremal" charge and entropy is, by equation (\ref{K}),
\begin{equation}\label{BETA}
\m{T_{NE}\;=\;{1\over 2\times 3^{1/4}\times \pi^{3/2}}\,\Bigg[{M\over K^3L^6}\Bigg]^{1/4}}.
\end{equation}
From Figure 2 we see at once that this sets a lower bound to the temperature of the black hole: we must have
\begin{equation}\label{GAMMA}
\m{T_{}\;\geq \;T_{NE}\;\approx\;0.068228\times \Bigg[{M\over K^3L^6}\Bigg]^{1/4}}.
\end{equation}
This is the result we seek. If one steadily increases the charge of the black hole, its entropy never reaches zero, but \emph{nor does its temperature}: the process effectively destabilizes the black hole before zero temperature can be reached. The situation can be encapsulated in the following remarkably simple formula: the minimal possible values of the entropy and the temperature [which we re-label accordingly], given in equations (\ref{Z}) and (\ref{BETA}) , satisfy
\begin{equation}\label{DELTA}
\m{S_{min}T_{min}\;=\;{1\over 3}M}.
\end{equation}
The absence of the geometric parameters K and L from the right hand side is striking; black holes of mass M in asymptotically locally AdS spacetimes, with flat conformal infinities, of whatever spatial periodicity, all have the same relationship between minimal entropy and minimal temperature as their charge varies. A computation of the minimal black hole entropy from an analysis of the microscopic degrees of freedom would therefore allow an actual evaluation of the minimal temperature, without requiring any knowledge of the geometric parameters.

As we discussed earlier, a lower bound on the temperature implies the existence of upper bounds on the chemical potential $\mu$ and on its temperature-normalized version,
$\bar{\mu}$. Combining equations (\ref{M}), (\ref{W}), and (\ref{Z}), we obtain
\begin{equation}\label{EPSILON}
\m{\mu_{NE}\;=\;{3^{1/4}\over 4 \pi^2}\,\Bigg[{M\over K^3L^6}\Bigg]^{1/4}},
\end{equation}
so we have
\begin{equation}\label{ZETA}
\m{\mu \; \leq \;\mu_{NE}\;\approx\;0.033337 \times \Bigg[{M\over K^3L^6}\Bigg]^{1/4}},
\end{equation}
which can be compared with the inequality in (\ref{N}).

The case of $\bar{\mu}$ is much more striking: we find that the upper bound, which we re-label appropriately, is simply
\begin{equation}\label{THETA}
\m{\bar{\mu}_{max}\;=\;\mu_{NE}/T_{NE}\;=\;\sqrt{{3\over 4 \pi}}},
\end{equation}
so that
\begin{equation}\label{IOTA}
\m{\bar{\mu}\;\leq \;\bar{\mu}_{max}\;\approx\;0.488603},
\end{equation}
contrasting sharply with the situation summarized in (\ref{NNN}). As in (\ref{DELTA}), the absence of K and L is remarkable.

As is customary, we must remind the reader that the field theory in question here is not QCD; one hopes that some kind of universality might render these results relevant to the experimental situation revealed by the celebrated RHIC experiment. In view of the extreme simplicity of our methods and the many assumptions made, even order-of-magnitude agreement is not to be looked for. With these disclaimers, we nevertheless give the following very brief and over-simplified discussion of the experimental situation.

As is well known, the RHIC experiment is believed to have produced a high-energy-density medium with fluid-like properties; the hope is that this medium has a dual description which is in some way similar to the ones we have been describing. One distinctive feature of this medium is that it is associated with a very high antiparticle/particle ratio, as compared with observations made in earlier experiments of a similar kind [but which did not produce such a medium]; see Figure 4 of \cite{kn:phobos}. A simple theoretical analysis shows that this ratio should be given by exp($-$2$\bar{\mu}_{\m{B}}$), where $\bar{\mu}_{\m{B}}$ is the temperature-normalized baryonic chemical potential. For RHIC collisions with a centre-of-mass energy of 200 GeV, the ratio is about 0.73, so that $\bar{\mu}_{\m{B}}$ is roughly 0.16. This of course agrees with (\ref{IOTA}). If we ignore the above disclaimers, then the prediction of (\ref{IOTA}) is that the antiproton/proton ratio cannot fall below about 0.38 in the fluid-like medium produced at the RHIC. Perhaps the best description of the situation is as follows: the systems observed at the RHIC have a relatively
high antiproton/proton ratio; we claim that holography predicts, in agreement with this, that there is in fact a lower bound on this ratio, if these systems are indeed dual to a [stable] black hole. This lower bound may not be very far below the observed value.

\addtocounter{section}{1}
\section* {\large{\textsf{6. Application: The Field Theory Phase Plane}}}
Let compare the situation we have found in Section 5 with the more familiar variety of black hole instability discussed in Section 3, above. There are two crucial differences.

First, the instability in Section 3 involved a transition of the black hole to another object with a well-defined geometry and correspondingly well-behaved dual physics, involving a confined phase of the field theory. The instability we have been discussing in Sections 4 and 5, by contrast, involves a transition to a state which is not well-defined, involving infinite reservoirs of negative free energy for branes. Thus, we do not have a transition to a confined state, but rather one to some kind of state which may perhaps have no dual geometric description at all; at any rate, if it does have a geometric dual, that dual is very different from the AdS soliton. It is natural,
in the specific case of the quark matter phase diagram, to associate this with the phase transitions from the QGP state to the ``CFL" and ``non-CFL" phases in Figure 1 of \cite{kn:alford}. Let us call these the ``exotic" phases.

Second, the instability in Section 3 gave rise to a lower bound on the temperature, but this lower bound depended only on K. Here, the temperature bound depends also on M, the mass of the black hole. A helpful way to think about this distinction is in terms of trajectories in the phase diagram, that is, the ($\mu$, T) plane. Suppose that we begin at some point in the region corresponding to the plasma phase: this corresponds to a specific choice of M and Q. Now gradually increase Q. The point in the phase diagram will move in the direction of decreasing T and increasing $\mu$, tracing out a curve which terminates on the phase boundary. Let us assume that this boundary represents the transition from the plasma to an exotic phase. Then we know that this point must satisfy $\bar{\mu}$ = $\sqrt{3/4\pi}$ = $\mu$/T; in other words, the boundary follows the straight line
\begin{equation}\label{IOTAIOTA}
\m{T\;=\;\sqrt{4\pi \over 3}\,\mu}.
\end{equation}
We saw in Section 3 that there is no black hole if the temperature falls below 1/(2$\pi$KL), so there is no occasion for brane-induced instability at low chemical potentials, since the straight line in (\ref{IOTAIOTA}) lies below 1/(2$\pi$KL) when $\mu$ is small. [Recall that the two soliton parameters A and B are constrained by only one relation, so we can readily arrange for the soliton to be immune to Seiberg-Witten instability: we just have to choose B to be small, and let A be fixed by K.] Thus the effect studied in Section 3 dominates at low values of $\mu$. It is only when the line in (\ref{IOTAIOTA}) intersects and rises above the horizontal line T = 1/(2$\pi$KL) that Seiberg-Witten instability comes to dominate, because then the cooling black hole becomes unstable in that way before it has an opportunity to make a transition to the soliton. That is, the instability discussed in Section 5 is the relevant one at high chemical potentials. In short, we have a transition, as the temperature decreases, to a confined phase at small $\mu$, but to some other, exotic phase when $\mu$ is large.

Thus we obtain a holographic version of the field theory phase diagram. As usual, we make no claim that this should be quantitatively accurate, but some qualitative features of the phenomenological phase diagram, which in any case is still very poorly understood, are reproduced by the holographic approach.

\addtocounter{section}{1}
\section* {\large{\textsf{7. Application: Limiting Violations of the KSS Bound}}}
The field theories we have been discussing are remote from QCD, so it came as a surprise when Kovtun et al. \cite{kn:son} were able to use such methods to make a remarkably accurate estimate [expressed in terms of the
 ``KSS bound"] for the viscosity/entropy density
ratio $\eta$/s of the medium observed at the RHIC. This estimate holds for a wide variety of systems, provided that one uses the standard two-derivative action for bulk gravity, as we have done throughout this work. However, in string theory one is interested in allowing for small corrections to this action, corrections involving higher-order derivatives: see \cite{kn:myers} and its references to earlier work. Unfortunately, the result is that the KSS bound appears to be violated. These violations are necessarily small, since the computation is itself based on the assumption that the couplings of the higher-order terms are suppressed by [powers of] the ratio of the Planck length to the AdS length scale. However, these small violations are enhanced by the inclusion of a chemical potential \cite{kn:myers}\cite{kn:cremonini}. The enhancement can be substantial if the temperature-normalized chemical potential can be very large; which, in fact, is the generic case in the absence of the effect described here. It is true that the value of $\bar{\mu}$ at the RHIC is small, so that the modification of the KSS result due to the chemical potential is likewise small there. But a quark plasma with large values of $\mu$ is potentially of physical interest [for example, in the study of the formation of neutron stars], so one wishes to know whether the KSS bound can fail badly in some region of the quark matter phase plane.

It is known that issues regarding causality and unitarity arise \cite{kn:ge2}\cite{kn:ge3}\cite{kn:ge4} when one goes beyond the Einstein-Hilbert gravitational lagrangian in the bulk, and it has been argued \cite{kn:brigante}\cite{kn:neupane}\cite{kn:sinha}, on these grounds, that $\eta$/s cannot be \emph{much} smaller than the value 1/4$\pi$ given by Kovtun et al.. In the context of the specific corrections discussed in \cite{kn:myers}\cite{kn:cremonini}, we can see that the effect discussed in this work also limits violations of the KSS bound: the temperature of the black hole in the AdS bulk \emph{cannot} go low enough to produce large values of
$\bar{\mu}$ in the dual field theory. Let examine the consequences in detail. [We again assume throughout that, as Q varies, the field theory remains in the plasma state; that is, that the black hole does not relax to a soliton.]

Myers et al. \cite{kn:myers} show that the KSS ratio is given, if a four-derivative action is used, by
\begin{equation}\label{KAPPA}
\m{{\eta\over s}\;=\;{1\over 4\pi}\,\Bigg[1\;-\;8c_1 \;+\;{16\bar{\mu}^2\over 3\Big(1\;+\;\sqrt{1\;+\;2\bar{\mu}^2/3}\Big)^2}\,\times\,\Big(c_1\;+\;6c_2\Big)\Bigg]}.
\end{equation}
Here c$_1$ is the dimensionless coupling of the squared curvature term $\m{R_{abcd}R^{abcd}}$, and c$_2$ is the dimensionless coupling of a term of the form $\m{R_{abcd}F^{ab}F^{cd}}$, where F$\m{_{ab}}$ is the field strength tensor; we remind the reader that, throughout this work, we have adopted the same normalization of the chemical potential as in \cite{kn:myers}. The addition of such terms to the action will of course modify our discussion in this work [for example, the precise width of the range of sub-extremal charges for which the black hole is unstable will change slightly, in a way that depends on these and other couplings], but, since all of these couplings must be extremely small, it is consistent to ignore this in a perturbative analysis.

Myers et al. give explicit examples of specific theories in which both correction terms in (\ref{KAPPA}) are negative,
so the KSS bound is violated here, and, since the function of $\bar{\mu}$ in the second correction term is an increasing function, large values of the normalized chemical potential do indeed make the situation significantly worse. In fact, this function is asymptotic to the value 8 at large values of $\bar{\mu}$. But our bound (\ref{IOTA}) forces it to be very much smaller: it can in fact be no larger than approximately
0.295249. If we assume that c$_1\;+\;$6c$_2$ is negative, then this means that we have a lower bound on the KSS ratio:
\begin{equation}\label{LAMBDA}
\m{{\eta\over s}\;\geq\;{1\over 4\pi}\Big[1\; -\; 7.704751\,c_1\;+\;1.771493\,c_2 \Big]}.
\end{equation}
In the case of the ungauged $N$ = 2 supergravity [for which c$_1$ is positive, and c$_2 = -\,$c$_1$/2], this simplifies to
\begin{equation}\label{MU}
\m{{\eta\over s}\;\geq\;{1\over 4\pi}\Big[1\; -\; 8.590498
\,c_1 \Big]}.
\end{equation}
Comparing this with equation (\ref{KAPPA}) when $\bar{\mu}$ = 0, we see that, even in the most extreme case,
the chemical potential can worsen the violation of the KSS bound only to a small extent.

To summarize: since c$_1$ and c$_2$ are small, and the effect of the chemical potential is marginal, we see that the
inclusion of higher-derivative corrections in the bulk action probably does not lead to large violations of the KSS
bound. A more precise formulation of this statement would of course be welcome.

\addtocounter{section}{1}
\section* {\large{\textsf{8. Conclusion}}}
Our principal conclusion is very simple: for one reason or another ---$\,$ and the two reasons are indeed very different ---$\,$ a black hole with a flat event horizon cannot be arbitrarily cold. Holographically, this means that the same statement holds true of the plasma phase of the field theory. This may seem obvious, but it is not true classically. Among other applications, this result implies that there is a surprisingly severe restriction on the range of possible temperature-normalized chemical potentials, given by (\ref{IOTA}). While the numerical value we found should not be taken too literally, it is not enormously different from the value of the baryonic temperature-normalized chemical potential in the fluid-like medium observed at the RHIC.

The instabilities responsible for all of our results are solely due to subtle properties of black hole thermodynamics and non-perturbative string theory; no such effects are seen otherwise \cite{kn:konoplya}. On the other hand, the requirement that the bulk should be free of Seiberg-Witten instability is undoubtedly just one of many consistency conditions imposed by
string theory; certainly other, completely different forms of string-related instability are known ---$\,$ see for example \cite{kn:gubsermit}\cite{kn:yamada}\cite{kn:horpolch}. In view of our findings here, it seems reasonable to hope, for example, that a full understanding of the KSS bound will be realised when all of these effects are taken into account. One should also attempt to understand the various ways in which still more general black holes,
involving other kinds of bulk matter, might fail to be stable, as for example in \cite{kn:singnojodi}. Again, such instabilities might, in the analogous asymptotically AdS contexts, impose interesting restrictions on a dual field theory defined on a flat spacetime.

At high values of the chemical potential, our results mean that even very hot black holes can be unstable. We understand this result on the gravitational side of the duality, but one would also like to see precisely what happens in the field theory in these circumstances: that is, to pinpoint the field theory analogue of brane instability in the bulk. One understands \cite{kn:seiberg} why there is an instability in the field theory when the scalar curvature at infinity is negative ---$\,$ the scalar curvature acts like a negative squared mass. But the situation is less clear when the scalar curvature at infinity is zero; some higher-order effect is responsible for the instability. An analysis of this point could be very useful.

In the application to the strongly coupled QGP \cite{kn:valdivia}, the lower bound on the temperatures of ``flat" black holes imposed by requiring stability in the Seiberg-Witten sense has a natural dual interpretation: there must always be a phase transition, to some ``exotic" state, if one tries to lower the temperature of the QGP at very high chemical potentials. This is of course precisely how the quark matter phase diagram is usually drawn \cite{kn:alford}. The question now is: can one use the holographic version of the phase diagram to give rough estimates of the temperatures and chemical potentials at which phase changes are to be expected, as future experiments scan across the diagram? This seems very ambitious, but, once again, the concept of ``black hole universality" may help us here.

\addtocounter{section}{1}
\section*{\large{\textsf{Acknowledgements}}}
The author is grateful to Dr. Soon Wanmei for helpful discussions and to Jude L$\ddot{\m{u}}$wen McInnes for encouraging him to complete this work expeditiously.

\end{document}